\documentstyle[prd,eqsecnum,aps,preprint,tighten,amssymb]{revtex}
\begin{document}
\draft
\title{Elimination of resonant divergences from QED
in super-strong magnetic fields}
\author{Carlo Graziani\thanks{E-Mail Address: carlo@twinkie.gsfc.nasa.gov}}
\address{Laboratory for High-Energy Astrophysics, Code 665, NASA Goddard
Space Flight Center, Greenbelt, MD 20771}
\author{Alice K. Harding}
\address{Laboratory for High-Energy Astrophysics, Code 665, NASA Goddard
Space Flight Center, Greenbelt, MD 20771}
\author{Ramin Sina}
\address{University of Maryland Department of Physics, College Park, MD 20742}
\date{November 15, 1994}
\maketitle
\begin{abstract}
We study the resonant divergences that occur in quantum scattering
cross-sections in the presence of a strong external magnetic field.  We
demonstrate that all such divergences may be eliminated by introducing
radiative corrections to the leading-order scattering amplitudes.
These corrections impose a choice of basis states that must be used in
scattering calculations: electron states must diagonalize the mass
operator, while photon states must diagonalize the polarization
operator.  The radiative corrections introduce natural line-widths into
the energy denominators of all propagators, as well as into the
time-development exponentials of all scattering states corresponding
to external lines.  Since initial and final scattering states may now
decay, it is logically necessary to compute scattering amplitudes for a
finite time-lapse between the preparation of the initial state and the
measurement of the final state.  Strict energy conservation, which
appeared in previous formulations of the theory, must thus be abandoned.
We exhibit the generic formulae for the scattering cross-sections in two
useful limits, corresponding to the cases where either the initial states
or the final states are stable, and discuss the application of the
general formula when neither of these limits applies.
\end{abstract}
\section{Introduction}
Astrophysicists have had a long-standing interest in the physics of
elementary processes in super-strong magnetic fields, with field
strengths $B\gtrsim 10^{12}$~G.  The cyclotron lines observed in the
spectra of Her X-1 \cite{tr78} and of 4U 0115+63 \cite{gr80} as well as
in many other X-ray pulsars have energy centers which correspond to
field intensities in this range.  There is also evidence for such field
strengths in the spin-down rates of radio pulsars.  If the spin-down is
attributed to energy loss to electromagnetic radiation from a spinning
magnetic dipole, many observations are consistent with field strengths
of the order of $10^{12}$-$10^{13}$~G, with some pulsar field strengths
well in excess of even $10^{13}$~G \cite{ha91,ts86}.  In addition,
there is tantalizing evidence for cyclotron lines in the spectra of
gamma-ray bursts seen by the gamma-ray burst detector aboard the GINGA
satellite \cite{mu88}, with line center energies consistent with field
intensities of order $10^{12}$~G.

The association of Soft Gamma Repeaters with supernova remnants
provides indirect evidence for even stronger fields.  If the 8~s
periodicity of the March 5 1979 event is identified with the rotation
period of the neutron star, the known age of the N49 remnant may be
used to estimate that the field strength is approximately $6\times
10^{14}$~G \cite{dt92}.  Moreover, such a strong field could help
resolve the puzzle of how the March 5 1979 event could ostensibly have
been so extravagantly in excess of the Eddington limit ($L\approx
10^4L_{\text{Edd}}$), by suppressing the Thomson cross-section for
photons propagating nearly parallel to the field lines \cite{pa92}.

In fields such as these, comparable in strength to the critical field
strength $B_c\equiv m^2c^3/e\hbar = 4.414\times 10^{13}$~G, all
calculations of elementary processes must be carried out using Quantum
Electrodynamics.  There have been many such calculations over the past
two decades, covering topics such as cyclotron absorption \cite{dv78},
cyclotron decay \cite{mz81,hrw82}, single photon pair production
\cite{k54,dh83}, pair annihilation to a single photon
\cite{w79,db80,h86,wphr86}, Compton scattering
\cite{h79,dh86,bam86,hd91}, two photon pair production \cite{km86}, two
photon pair annihilation \cite{w79,db80}, $\text{e}^-\text{e}^-$
scattering \cite{l81}, and several more.  All these processes have very
different behavior from their $B=0$ counterparts (if those counterparts
are even possible), on account of the peculiar kinematics,as well as
the discrete electronic states (Landau levels) associated with a
uniform external magnetic field.

These calculations have always been carried out in the Furry picture,
in close analogy to the $B=0$ Feynman rules.  The free space electron
propagator is replaced by a propagator which is a Green's function for the
Dirac equation in the external field, and the external fermion lines
are represented by solutions of that equation.  The results have often
been interesting and useful, but they have not been uniformly
satisfactory.  The leading order calculation of resonant Compton
scattering yields results which are divergent at the cyclotron
resonances \cite{h79,dh86}, evidently because to this order the theory
makes no provision for natural line width.  The line width may be
included ``by hand'' in the results, making the cyclotron resonances
finite \cite{bam86,hd91}.  However, there are other such ``resonant
divergences'' in Compton scattering, which have nothing to do with the
cyclotron resonances and which are not nearly as tractable
\cite{h79,dh86}.  In fact, one such resonance, which occurs exactly
at the threshold where the initial photon may pair-produce, is responsible
for making the total Compton cross-section divergent {\em everywhere}
above this threshold.

In fact, the theory of elementary processes in external magnetic fields
is plagued with such divergences.  With a little practice, it is not
hard to discover divergences resembling these {\em in every single
process of second or higher order}.  Clearly, this is a troublesome
development that casts a shadow upon the entire undertaking.

These resonant divergences occur because the kinematics of these
processes allow intermediate ``virtual states'' to be real --- that is,
on-shell.  Often, the on-shell intermediate state is an excited state,
such as an electron in an excited Landau level, or a photon above the
one photon pair-production threshold.  In this case there is an
associated decay width that may be pressed into service to control
the divergence.  The Weisskopf-Wigner broadening prescription,
\begin{equation}
E\rightarrow E-\frac{1}{2}i\Gamma,\label{ecmplx}
\end{equation}
when applied to the energy denominators in the propagator, pushes the
poles in the propagator off the real axis, so that while the
intermediate states may still be on-shell the propagator no longer
diverges there.  This is in fact the approach that has been adopted for
Compton scattering \cite{bam86,hd91,g93}, and which ascribes the
natural line width to the cyclotron resonances.

However, there are circumstances in which a {\em stable} on-shell
intermediate state may be produced.  Such states are not attended by a
decay width, so the associated divergence may not be reined in as
before.

It should be pointed out that these divergences are entirely unrelated
to the notorious ultraviolet divergences of QED.  They
are not the consequence of improper manipulation of field-theoretic
distributions; rather, they occur whenever the circumstances of the
elementary process permit a kinematically accessible on-shell
intermediate state (KAOSIS).  It is easy to recognize when a KAOSIS is
permitted.  For example, a second-order process will allow one if it
may be viewed as a succession of two real first-order processes.
Thus the KAOSIS corresponding to cyclotron resonance occurs because
the process may be viewed as a real cyclotron absorption followed
by a real cyclotron emission.  Similarly, the second ``disastrous''
resonance in Compton scattering is due to a KAOSIS that corresponds
to the initial photon undergoing a real decay to a pair, followed
by the resulting positron annihilating with the initial electron
to produce the final photon.  The reason this KAOSIS is catastrophic
is that the intermediate positron may be in the Landau ground state,
so that no decay width is available to restrain the divergence.

At the same time, there is a second defect of the theory which so far
has not received recognition as a problem.  The calculation of S-matrix
elements as outlined above always results in a $\delta$-function which
enforces strict energy conservation between initial and final states.
This remains true even if some of these scattering states are
unstable.  But this is not physically sensible;  the energy of an
unstable state is only known to within its decay width, so that it is
ludicrous to demand strict energy conservation for such transitions.
Nevertheless, the current theory does so irrespective of whether or not
the states are stable.  As an example, the calculation of the decay of
an electron in an excited Landau state produces the result that the
emitted cyclotron photon is monochromatic, rather than having the
Lorentzian line shape characteristic of resonant decay
\cite{mz81,hrw82}.

The difficulties of resonant divergence and of spurious energy conservation
are related.  Briefly, a stable KAOSIS may only occur if some of the
particles in the initial and final states are themselves unstable, and
it is their decay widths that restrain the divergence.  However, the
introduction of these decay widths smears out their energy, so that
energy conservation (which was a consequence of their assumed eternal
duration) no longer obtains.

Thus it appears that we must modify the current theory somehow if we
are to circumvent these unphysical features.  That modification is the
purpose of this work.  We demonstrate below that radiative corrections
modify the propagators and the scattering states by introducing their
respective decay widths into the S-matrix elements.  In this sense, we
are extending the results of Graziani \cite{g93}, who carried out this
program for the electron propagator only.  The corrected ``dressed''
states and propagators result in scattering cross-sections that are
always finite.

In Sec.\ \ref{sec_rdp} we discuss the role of the bare propagators
in producing resonant divergences.  In Sec.\ \ref{sec_dep} we review
the work of reference \cite{g93} on the electron propagator, which
we extend to the photon propagator in Sec.\ \ref{sec_dpp}.  We
exhibit the corrections to the scattering states in Sec.\ \ref{sec_css}.
In Sec.\ \ref{sec_sf} we derive the modification of the S-matrix elements,
and exhibit two useful limits: the absorption limit and the emission limit,
wherein the initial and final states are stable, respectively.  Finally,
in Sec.\ \ref{sec_disc} we discuss the applicability of these results
and the extent of the modification from the ``standard'' theory.

\section{Resonant divergences and propagators}\label{sec_rdp}
We begin with a discussion of the role that the bare electron and
photon propagators have in producing the resonant divergences.  In this
section and throughout the rest of this work we use the metric
signature $[+1,-1,-1,-1]$.

The bare electron propagator is simply expressed in terms of the
fermion one-particle states.  We work in the Furry picture, so these
states are represented by solutions $\Psi^{(\eta)}_A$ of the
Landau-Dirac (L-D) equation, that is, the Dirac equation in the
presence of a classical, uniform, external magnetic field:
\begin{equation}
[\gamma^\mu(i\partial_\mu-eA_\mu)-m]\Psi^{(\eta)}_A=0.\label{ldeqn}
\end{equation}
Here, $\eta=\pm 1$ refers to whether the solution has positive or
negative energy, while $A$ is the set of quantum numbers that specify
the state --- Landau level number, $x^3$-momentum, spin, and an orbit
center coordinate.  We choose the gauge $A^0=0$, ${\bf A}=Bx^1{\bf
e}_2$ for the external magnetic field, which is directed in the ${\bf
e}_3$ direction.  We also choose the orbit center coordinate $a$ to be
the $x^1$ Cartesian coordinate of the orbit center.  With these
choices, the functional dependence of the solutions of
Eq.\ (\ref{ldeqn}) is that of a plane wave in the $x^2$ and $x^3$
coordinates, while the $x^1$ dependence is that of a one-dimensional
harmonic oscillator eigenfunction centered about $x^1=a$ \cite{jl49,mp83}.

It is convenient to separate out the time dependence of $\Psi^{(\eta)}_A$:
\begin{equation}
\Psi^{(\eta)}_A(x)=e^{-i\eta E_Ax^0}\phi^{(\eta)}_A({\bf x}),\label{esep}
\end{equation}
where $\phi^{(\eta)}_A({\bf x})$ is an eigenstate of the Landau-Dirac
Hamiltonian with quantum numbers $A$ and eigenvalue $E_A$.  $E_A$ is
given by $E_A=[m^2+(p^3)^2+2mN\omega_B]^{1/2}$, where $N$ is the Landau
level number, $p^3$ is the momentum in the $x^3$ direction, and
$\omega_B$ is the Larmor frequency.  The bare electron propagator
$G_B(x,y)$ may be represented in terms of the $\phi^{(\eta)}_A$:
\begin{equation}
G_B(x,y)={\displaystyle\sum_{A,\eta}}\int\frac{dp^0}{2\pi}\,e^{-ip^0(x^0-y^0)}
\frac{\phi^{(\eta)}_A({\bf x})\overline{\phi^{(\eta)}_A}({\bf y})}
{p^0-\eta(E_A-i0)}.\label{beprop}
\end{equation}
It is easy to verify that the expression in Eq.\ (\ref{beprop}) is a
Green's functions for Eq.\ (\ref{ldeqn}) with the required boundary
conditions \cite{bd64}.  We write $G_B(x,y)$ rather than $G_B(x-y)$
because neither Eq.\ (\ref{ldeqn}) nor its Green's functions are
translationally invariant --- they are invariant under combinations of
translations and gauge transformations.

The bare electron propagator enters the leading order calculation of
such processes as Compton scattering, pair annihilation to two photons,
and two photon pair production.  The S-matrix elements for these
processes are obtained by sandwiching $G_B(x,y)$ between appropriate
fermion and photon states and integrating over $x$ and $y$.  These
processes all exhibit resonant divergences.  The proximate cause of
those divergences is evidently the energy denominator in
Eq.\ (\ref{beprop}).  The convolution of the propagator with the
initial and final states fixes the values of $p^0$ and $p^3$ at the
total energy and $x^3$-momentum of the states, respectively.  Thus, the
residual
energy-related degree of freedom in the summation over intermediate
states is the Landau level number $N$, a non-negative integer.  It follows
that the energy denominator in Eq.\ (\ref{beprop}) may be zero if the
energy and $x^3$-momentum of the scattering states are suitably tuned
to one of the Landau levels.  If this happens a resonant divergence occurs.

The bare photon propagator has the usual ``transverse'' expression
\begin{equation}
D_B(x-y)^{\mu\nu}=\int\frac{d^4k}{(2\pi)^4}\,(-g^{\mu\nu}+k^\mu k^\nu/k^2)
\frac{e^{-ik(x-y)}}{k^2+i0}.\label{bpprop}
\end{equation}
It enters the calculation of S-matrix elements for processes such as
electron-electron scattering \cite{l81} and electron-positron
scattering.  These processes have total cross-sections which exhibit
resonant divergences.  This may seem surprising at first, since the
S-matrix elements for these processes are entirely finite, even when
a KAOSIS is present.  There is a key difference between the photon and
electron propagators:  In the electron case the on-shell energy
contains a dependence on the Landau level quantum number, a discrete
degree of freedom.  Thus the resonances are ``spaced-out'' and
well-separated.  On the other hand, the photon on-shell energy is
entirely dependent on the purely continuous degrees of freedom ${\bf
k}$, and the ``sum over intermediate states'' is really an integral.
While this integral will undoubtedly encounter any KAOSIS singularity,
the small imaginary part in the denominator of the propagator provides
a perfectly straightforward prescription for circumventing the pole.
Thus the S-matrix elements are finite for these processes.
Nevertheless, whenever an on-shell intermediate state is kinematically
accessible, the total cross-sections diverge.  What is going on?

The divergence is evident in the expressions in Appendix C of Langer
\cite{l81} for the e-e scattering cross-section.  Langer performed
the integrals over the intermediate states simultaneously with
those over final states and the average over initial states, in
order to take advantage of several simplifications that ensue.
Unfortunately, these manipulations obscure the source of the divergences.
In order to shed some light on the situation, we have calculated
cross-sections for e-e and e$^+$-e$^-$ scattering by computing
the S-matrix elements before summing over final and averaging over
initial states.

What we have found can be best illustrated by considering the specific
example of e$^+$-e$^-$ scattering.  As depicted in Fig.\ \ref{e+e-},
there are two relevant Feynman diagrams.  We have confirmed that in
each case the S-matrix elements are finite even when there is a KAOSIS.
When a KAOSIS exists, however, there is a divergence resulting from the
sum over {\em final} states.  In the diagram of Fig.\ \ref{e+e-}(a),
the divergence arises from the integral over the variable
$s=a_f^+-a_f^-$, where $a_f^+$ and $a_f^-$ are
the orbit center $x^1$-coordinates of the final positron and electron,
respectively.  In the diagram of Fig.\ \ref{e+e-}(b), the divergence
is due to the integral over the variable $t=a_f^--a_i^-$, where
$a_i^-$ is the orbit center $x^1$-coordinate of the initial electron.

We see, then, that the divergence manifests itself as an infinite range
of interaction.  When a KAOSIS is present, the intermediate photon may
travel an arbitrarily large distance between its emission and its
absorption without a reduction in amplitude.  Consequently there is no
spatial cutoff in the interaction, and the resulting integral over
orbit center separation diverges.

The situation may be further clarified by explicitly computing the
interaction range in the $x^1$ direction.  When we calculate an
S-matrix element for these processes, we integrate the bare photon
propagator $D_B(x-y)^{\mu\nu}$ multiplied by two fermion currents, one
for each vertex.  Since our states are chosen to be plane waves in
the $x^2$-$x^3$ direction, the currents are also plane waves in
$x^2$ and $x^3$, and the integrals over $x^0$, $x^2$, and $x^3$
are tantamount to Fourier transforms of the photon propagator in
those coordinates.  The remaining integral over $x^1$ folds
together two (in general well-separated) current functions
$j_x(x^1)^\mu$, $j_y(y^1)^\nu$ with the transformed propagator
$f_B(x^1-y^1;k^0,k^2,k^3)g_{\mu\nu}$.  We may easily compute
$f_B(x^1;k^0,k^2,k^3)$.  If we define
$\rho\equiv (k^0)^2-(k^2)^2-(k^3)^2=\pm q^2$, with $q\ge 0$, we find
\begin{equation}
f_B(x^1;k^0,k^2,k^3)=\left\{\begin{array}{ll}
-\frac{i}{2q}e^{iq|x^1|} & \mbox{if $\rho\ge 0$,}\\
-\frac{1}{2q}e^{-q|x^1|} & \mbox{otherwise.}
\end{array}\right.\label{effrange1}
\end{equation}
Thus, if $\rho\ge 0$ (the condition for the existence of a KAOSIS), the
effective interaction range is infinite, whereas it is of finite range
$\sim 1/q$ otherwise.

The situation is reminiscent of Coulomb scattering.  If a wave packet
scatters with large impact parameter from the center of the Coulomb
potential, the momentum transfer is very low, and thus the exchanged
photon is very close to its mass-shell (that is, its $k$ vector is very
near the apex of the light cone).  This is the reason that the total
Coulomb cross-section is divergent --- the extreme infrared photons,
which are nearly on-shell, give the interaction an infinite range.
There is one noteworthy difference here, however:  for the processes we
consider, the virtual photons do not need to be infinitely soft to be
on-shell, so that these resonant divergences are in no sense infra-red
divergences.

As a consequence of this analysis, the cure for
photon-propagator-related resonant divergences will necessarily have a
slightly different mathematical character from that for the divergences
that arise in connection with the electron propagator.  While in the
latter case we need to show how the previously diverging amplitudes may
be made to converge, in the former case we must show how the previously
long-range ``effective interaction'' may have its range curtailed.

\section{The dressed electron propagator}\label{sec_dep}
The necessity of complexifying the energy denominator of the electron
propagator so as to obtain the Weisskopf-Wigner line shape has been
recognized for some time \cite{bam86,hd91}, although these authors
lacked a formal justification for their {\em ad hoc} complexification
of the energy, given in Eq. (\ref{ecmplx}).
Graziani\cite{g93} provided this justification by considering radiative
corrections to the electron propagator.  As it turns out, the same
procedure may be applied to the photon propagator as well.  Quite
generally, one finds that the operator which represents the
self-interaction (the mass operator $\Sigma$ in the case of the
electron, the polarization operator $\Pi$ in the case of the photon)
singles out those solutions of the wave equation in which it is
diagonal.  The dressed propagator may then be expressed as a sum over
all such states, with the energy in the denominator acquiring an
imaginary part given by Eq.\ (\ref{ecmplx}).  We review the discussion
of the electron propagator from Ref.\ \cite{g93}, and extend it to the
photon propagator in the next section.

The dressed propagator $G(x,y)$ is related to the bare propagator and
the self-energy operator $\Sigma(x,y)$ by the Dyson equation
$G=G_B+G_B\cdot\Sigma\cdot G$, where the dot denotes four-dimensional
convolution as well as spinorial matrix multiplication.  It follows
that $G(x,y)$ is a Green's function for the dressed L-D operator

\begin{equation}
[\gamma^\mu(i\partial_\mu-eA_\mu)-m-\Sigma]\cdot\Psi^{(\eta)}_A=0.
\label{dldeqn}
\end{equation}

Now, let $\Theta^{(\eta)}_{p^0A}(x)=e^{-ip^0x^0}\phi^{(\eta)}_A({\bf
x})$.  It may be shown \cite{mp83,p87,g93} that $\Theta$ diagonalizes
$\Sigma$ if the $\phi^{(\eta)}_A$ are simultaneous eigenstates of the
Landau-Dirac Hamiltonian and of the $x^3$-component of the magnetic
moment operator of Sokolov and Ternov \cite{st68}.  This condition
defines what is meant by ``spin up'' and ``spin down'', in the absence
of the tools provided by Poincar\'e group representation theory (the
physical system no longer has full Poincar\'e invariance).  If this
condition is satisfied, {\em and only in this case}, $G$ may be
represented in a form analogous to Eq.\ (\ref{beprop}):
\begin{equation}
G(x,y)={\displaystyle\sum_{A,\eta}}\int\frac{dp^0}{2\pi}\,e^{-ip^0(x^0-y^0)}
\frac{\phi^{(\eta)}_A({\bf x})\overline{\phi^{(\eta)}_A}({\bf y})}
{p^0-\eta(E_A-i0)-\Sigma(p^0,A,\eta)},\label{dp1}
\end{equation}
where $\Sigma(p^0,A,\eta)$ is the diagonal matrix element of $\Sigma$
in the state $\Theta^{(\eta)}_{p^0A}$.  It is in general a complex
number, in contradistinction to the case where the external field
strength is zero, where the diagonal matrix elements of the self-energy
operator are real.  The real part of $\Sigma(p^0,A,\eta)$ merely yields
a small shift in the energy of the state, while the imaginary part
provides a line-width to the otherwise divergent resonant energy
denominator.  This justifies neglecting the real shift while preserving
the imaginary width.  It was shown explicitly in Ref.\ \cite{g93} that
in this approximation, the pole in the propagator corresponding to
state $A$, $\eta$ is located at $p^0=\eta(E_A-\frac{i}{2}\Gamma_A)$,
where $\Gamma_A$ is just the Weisskopf-Wigner decay rate of the state
$A$, computed to leading order:
\begin{equation}
\Gamma_A=e^2\int\frac{d^3{\bf k}}{2\omega_k(2\pi)^3}\,
{\displaystyle\sum_{\epsilon}\sum_B}
|T_{AB}({\bf k},\mbox{\boldmath$\epsilon$})|^2\,(2\pi)\,
\delta(E_A-E_B-\omega_k).\label{gammadef}
\end{equation}
Here, $T_{AB}({\bf k},\mbox{\boldmath$\epsilon$})$ is the interaction
matrix element for the transition from the state $A$ to the state $B$
with the emission of a photon with wave vector {\bf k} and polarization
\mbox{\boldmath$\epsilon$}.
It follows that the dressed propagator may be expressed as
\begin{eqnarray}
G(x,y)&=&{\displaystyle\sum_{A,\eta}}\int\frac{dp^0}{2\pi}\,e^{-ip^0(x^0-y^0)}
\frac{\phi^{(\eta)}_A({\bf x})\overline{\phi^{(\eta)}_A}({\bf y})}
{p^0-\eta(E_A-i\Gamma_A/2)}\label{depropa}\\
&=&-i{\displaystyle\sum_{A,\eta}}\eta\,\theta\left[\eta\,(x^0-y^0)\right]
e^{-i\eta (E_A-i\Gamma_A/2)(x^0-y^0)}
\phi^{(\eta)}_A({\bf x})\overline{\phi^{(\eta)}_A}({\bf y})\label{depropb}
\end{eqnarray}
to first order in $e^2$.

It appears from these equations that the prescription of Eq.\ (\ref{ecmplx})
is in fact rigorously justifiable.  We wish to emphasize, however, that
Eqs.\ (\ref{depropa})-(\ref{depropb}) are only correct when the states
$\phi^{(\eta)}_A$ are chosen so that the $\Theta^{(\eta)}_{p^0A}$ diagonalize
the self-energy operator, or equivalently, so that the $\phi^{(\eta)}_A$
diagonalize the $x^3$-component of the Sokolov-Ternov magnetic moment operator.
The states of Sokolov and Ternov \cite{st68}, and those of Herold, Ruder,
and Wunner \cite{hrw82} satisfy these conditions, and their use leads to
correct expressions for the scattering cross-sections.  The states of Johnson
and Lippmann \cite{jl49}, which have gained some currency in the literature,
do not satisfy the required conditions \cite{mp83}.  In particular, as
discussed in \cite{g93}, the use of Johnson-Lippmann states in the
computation of cyclotron scattering cross-sections can lead to relative
errors of order 45\% at the first cyclotron harmonic, depending on the field
strength.
\section{The dressed photon propagator}\label{sec_dpp}
The procedure for the photon propagator is entirely analogous to the
one followed for the electron propagator.  The self-interaction of the
Maxwell field is represented by the polarization operator
$\Pi(x-y)^{\mu\nu}$, which satisfies the condition of gauge invariance,
$\frac{\partial}{\partial x^\mu}\Pi(x-y)^{\mu\nu}=0$.  The dressed
photon propagator $D(x-y)^{\mu\nu}$ is obtained from the bare
propagator $D_B(x-y)^{\mu\nu}$ and $\Pi(x-y)^{\mu\nu}$ by solving the
Dyson equation $D=D_B+D_B\cdot\Pi\cdot D$.  In order to accomplish
this, it is necessary to find the polarization states which diagonalize
$\Pi^{\mu\nu}$, which are analogous to the spinor states that
diagonalize $\Sigma$.  These polarization states were found by Batalin
and Shabad \cite{bs71}, and written explicitly for the case of a
uniform magnetic field by Shabad \cite{s75}.  For a photon with
4-momentum $k$ propagating in a uniform magnetic field ${\bf B}=B{\bf
e}_3$ we have the following three (unnormalized) polarization vectors:
\begin{mathletters}
\label{poldef}
\begin{equation}
b_\perp^\mu=k^2(e_1)^\mu-k^1(e_2)^\mu,\label{polperp}
\end{equation}
\begin{equation}
b_\|^\mu=k^3(e_0)^\mu+k^0(e_3)^\mu,\label{polpar}
\end{equation}
\begin{equation}
b_{\text L}^\mu=(k^\nu k_\nu){k_\perp}^\mu-({k_\perp}^\nu{k_\perp}_\nu)k^\mu,
\label{pollong}
\end{equation}
\end{mathletters}
where ${k_\perp}^\mu=k^1(e_1)^\mu+k^2(e_2)^\mu$, and $(e_\rho)^\mu$ is
a unit vector in the $x^\rho$ direction.  These modes diagonalize the
Fourier-transformed polarization operator $\Pi(k)^{\mu\nu}=\int d^4x\,
\Pi(x)^{\mu\nu}e^{ikx}$.  There are three of them, because by gauge
invariance $\Pi$ satisfies $k_\mu\Pi(k)^{\mu\nu}=0$, so that the fourth
mode is just $k$, and the eigenvalue of $\Pi$ that corresponds to it is
zero.  The mode $b_\|$ is so labeled because on shell it differs from
the usual ``parallel'' polarization 3-vector by an inessential multiple
of $k$, while $b_\perp$ is just the usual ``perpendicular'' mode.  The
vector $b_{\text L}$ represents a longitudinal mode, which on shell is
proportional to $k$.  The $b$ in Eqs.\ (\ref{poldef}) are orthogonal to
each other, and to $k$.

Using these modes, the transverse photon propagator may be expressed
as follows:
\begin{equation}
D(x-y)^{\mu\nu}=-\int\frac{d^4k}{(2\pi)^4}\,e^{-ik(x-y)}
\displaystyle{\sum_{j}}\frac{b_j^\mu b_j^\nu}{b_j\cdot b_j}\,
\frac{1}{(k^0)^2-{\omega_k}^2+\Pi(k^0,{\bf k},j)},\label{dp2}
\end{equation}
where
\begin{equation}
\Pi(k^0,{\bf k},j)\equiv \Pi(k)_{\mu\nu}{b_j}^\mu{b_j}^\nu /(b_j\cdot b_j).
\label{pimel}
\end{equation}
To first order in $e^2$, the pole in Eq.\ (\ref{dp2}) is located at
$(k^0)^2={\omega_k}^2-\Pi(\omega_k,{\bf k},j)$, so that to this order
Eq.\ (\ref{dp2}) may be written
\begin{equation}
D(x-y)^{\mu\nu}=-\int\frac{d^4k}{(2\pi)^4}\,e^{-ik(x-y)}
\displaystyle{\sum_{j}}\frac{b_j^\mu b_j^\nu}{b_j\cdot b_j}\,
\frac{1}{(k^0)^2-{\omega_k}^2+\Pi(\omega_k,{\bf k},j)}.\label{dp3}
\end{equation}
Note that when the light-cone condition $k^0=\omega_k$ is satisfied,
both $b_\perp$ and $b_\|$ are space-like ($b\cdot b<0$) while the
longitudinal mode $b_L$ is light-like.  It follows that
$\Pi(\omega_k,{\bf k},{\text L})=0$, so we only need compute
$\Pi(\omega_k,{\bf k},j)$ for $j=\perp,\|$.

The (unrenormalized) leading-order expression for $\Pi(x-y)^{\mu\nu}$
is
\begin{equation}
\Pi(x-y)^{\mu\nu}=-ie^2\text{Tr}(G_B(y,x)\gamma^\mu G_B(x,y)\gamma^\nu).
\label{pidef}
\end{equation}
Following the analogy to the case of the electron propagator, we
calculate the imaginary part of the diagonal matrix elements of
$\Pi^{\mu\nu}$, while neglecting the real part.  For this purpose,
Eq.\ (\ref{pidef}) is entirely adequate, even though it is not
renormalized.  The renormalization counter-terms which are to be
subtracted from the diagonal matrix elements of $\Pi^{\mu\nu}$ are
purely real, so that the imaginary parts are unaffected by
renormalization.

This expression for $\Pi^{\mu\nu}$ is in fact translationally
invariant, even though $G_B$ is not.  The translational invariance
may be established using the ``translation+gauge transformation''
invariance of $G_B$ alluded to in the previous subsection \cite{ms76}.

Substituting Eq.\ (\ref{beprop}) into Eq.\ (\ref{pidef}), after
some manipulation we obtain
\begin{equation}
\Pi(x-y)^{\mu\nu}=-e^2{\displaystyle\sum_{A,A^\prime,\eta}}
\int\frac{dp^0}{2\pi}\,e^{-ip^0(x^0-y^0)}\,
\frac{[\overline{\phi^{(\eta)}_A}({\bf x})\gamma^\mu
\phi^{(-\eta)}_{A^\prime}({\bf x})]
[\overline{\phi^{(-\eta)}_{A^\prime}}({\bf y})\gamma^\nu
\phi^{(\eta)}_A({\bf y})]}
{\eta p^0+E_A+E_{A^\prime}-i0}.
\end{equation}
We now Fourier transform this equation.  Taking the implicit translation
invariance into account, we obtain
\begin{eqnarray}
\Pi(k)^{\mu\nu}\epsilon_\mu \epsilon_\nu&=&L^{-3}T^{-1}\int d^4x\, d^4y\,
e^{ik(x-y)}\Pi(x-y)^{\mu\nu}\epsilon_\mu \epsilon_\nu
\nonumber\\
&=&-2\omega_k{\displaystyle\sum_{A,A^\prime,\eta}}
\frac{|J_{AA^{\prime}}^{(\eta)}({\bf k})^\mu \epsilon_\mu|^2}
{\eta k^0+E_A+E_{A^\prime}-i0},\label{pimel1}
\end{eqnarray}
where
\begin{equation}
J_{AA^{\prime}}^{(\eta)}({\bf k})^\mu\equiv e^2L^{-3/2}(2\omega_k)^{-1/2}
\int d^3{\bf x}\,
\overline{\phi^{(\eta)}_A}({\bf x})
\gamma^\mu\phi^{(-\eta)}_{A^\prime}({\bf x})\,
e^{i{\bf k}\cdot{\bf x}}.\label{current}
\end{equation}
When $\epsilon^\mu$ is a normalized polarization vector,
$J_{A^{\prime}A}^{(+)}({\bf k})^\mu\epsilon_\mu$ is just the
interaction matrix element for a transition from a photon with wave
vector ${\bf k}$ and polarization {\boldmath$\epsilon$} to a pair with
quantum numbers $A^{\prime}A$.  The imaginary part of Eq.\ (\ref{pimel1})
is
\begin{equation}
\text{Im}[\Pi(k)^{\mu\nu}\epsilon_\mu \epsilon_\nu]=
-2\omega_k{\displaystyle\sum_{A,A^\prime}}
\left|J_{AA^{\prime}}^{(+)}({\bf k})^\mu \epsilon_\mu\right|^2\,
\pi\,\delta\left(E_A+E_{A^{\prime}}-|k^0|\right),\label{pimel2}
\end{equation}
where we have used the identity
$\sum_{AA^\prime}\left|J_{AA^{\prime}}^{(+)}({\bf k})^\mu
\epsilon_\mu\right|^2=
\sum_{AA^\prime}\left|J_{AA^{\prime}}^{(-)}({\bf k})^\mu
\epsilon_\mu\right|^2$,
a consequence of the parity invariance of Eq.\ (\ref{ldeqn}).

Comparing with Eq.\ (\ref{pimel}), we see that to obtain the imaginary
part of $\Pi(\omega_k,{\bf k},j)$, we may impose the light-cone
condition $k^0=\omega_k$ and let
$\epsilon_\mu=(\epsilon_j)_\mu=(b_j)_\mu/|b_j\cdot
b_j|^{1/2}\Big|_{k^0=\omega_k}$ in Eq.\ (\ref{pimel2}), keeping in mind
that $b_j\cdot b_j<0$.  The result is
\begin{eqnarray}
\text{Im}[\Pi(\omega_k,{\bf k},j)]&=&
2\omega_k{\displaystyle\sum_{A,A^\prime}}
\left|J_{AA^{\prime}}^{(+)}({\bf k})^\mu (\epsilon_j)_\mu\right|^2\,
\pi\,\delta\left(E_A+E_{A^{\prime}}-\omega_k\right)\nonumber\\
&=&2\omega_k\times\Gamma({\bf k},j)/2.\label{gammaphot}
\end{eqnarray}
Clearly, $\Gamma({\bf k},j)$ is just the Weisskopf-Wigner decay rate of
the photon state $({\bf k},j)$.  The energy denominator in Eq.\ (\ref{dp3})
is thus $(k^0)^2-{\omega_k}^2+2i\omega_k\Gamma({\bf k},j)/2\approx
(k^0)^2-(\omega_k-i\Gamma({\bf k},j)/2)^2$, to first order in $e^2$.
Consequently, the dressed photon propagator is
\begin{eqnarray}
D(x-y)^{\mu\nu}&=&-\int\frac{d^4k}{(2\pi)^4}\,e^{-ik(x-y)}
\displaystyle{\sum_{j}}\frac{b_j^\mu b_j^\nu}{b_j\cdot b_j}\,
\frac{1}{(k^0)^2-(\omega_k-i\Gamma({\bf k},j)/2)^2}.\label{dppropa}\\
&=&i\int\frac{d^3{\bf k}}{(2\pi)^32\omega_k}\nonumber\\
&&\quad\times\displaystyle{\sum_{j}\sum_{\eta=\pm 1}}\,
\frac{b_j^\mu b_j^\nu}{b_j\cdot b_j}\,
\theta\left[\eta\,(x^0-y^0)\right]\,
e^{-i\eta(\omega_k-i\Gamma({\bf k},j)/2)(x^0-y^0)}\,
e^{i\eta{\bf k}\cdot ({\bf x}-{\bf y})}.\label{dppropb}
\end{eqnarray}
We see that the prescription of Eq.\ (\ref{ecmplx}) continues to hold
in the case of the photon field.  It should be emphasized, however,
that it is essential that the polarization modes given in
Eqs.\ (\ref{poldef}) be used in the expression for the propagator in
order for that expression to be correct.

Recalling the discussion at the end of Sec.\ \ref{sec_rdp}, we
investigate the range of the ``dressed'' interaction by computing the
partial Fourier transform of the propagator in Eq.\ (\ref{dppropa})
with respect to $x^0$, $x^2$, and $x^3$.  Assuming the presence of
a KAOSIS (so that $q^2=(k^0)^2-(k^2)^2-(k^3)^2\ge 0$), we find for
the transformed propagator $f(x^1;k^0,k^2,k^3)$,
\begin{eqnarray}
f(x^1;k^0,k^2,k^3)&=&-\frac{i}{2Z(q)}e^{iZ(q)|x^1|},\nonumber\\
Z(q)&\equiv&\frac{1}{\sqrt{2}}\left(\sqrt{u^2+q^2}+i\sqrt{u^2-q^2}\right),
\nonumber\\
u^2&\equiv&\sqrt{q^4+(k^0)^2{\Gamma({\bf k},j)}^2},\label{effrange2}
\end{eqnarray}
where $\Gamma({\bf k},j)$ is evaluated at the point $k^1=q$.  Since the
imaginary part of $Z(q)$ is positive, we find that the presence of a
non-zero decay rate $\Gamma({\bf k},j)$ has cut off the interaction
range.

Note that $Z(q)\rightarrow q$ as $\Gamma\rightarrow 0$, so that if the
KAOSIS is below the one photon pair-production threshold the
interaction range is not cut off (in fact we then have $f=f_B$, as
expected).  In the next section, we will show how the interaction range
is cut off when the KAOSIS is below threshold.  In the mean time
though, we have already disposed of the resonant divergence associated
with the process of Fig.\ \ref{e+e-}(b), which always has a KAOSIS
above the one photon pair-production threshold.  Indeed, we may now
give a useful interpretation of that divergence: it arose because the
intermediate photon (which at the KAOSIS may be viewed as being due to
a pair annihilation) was not instructed to decay to a pair in a finite
time, so that it produced finite amplitude for the final pair at all
values of $x^1$.  The resulting total cross-section was infinite.  Now
that the dressed propagator is used to calculate the S-matrix element,
the photon is aware of its decay obligations, and the cross-section due
to this process is finite.

\section{Corrections to Scattering States}\label{sec_css}
The propagator corrections described in the previous section are
sufficient to control resonant divergences in many processes and
regimes.  Nevertheless, there are still cases where a scattering
process may lead to a divergent resonance.  Specifically, any process
exhibiting a KAOSIS with vanishing
decay rate will exhibit divergent scattering cross-sections even
if calculated using the dressed propagators.

Such processes are not at all rare.  Consider the case of an electron
scattering with a photon which is above the one photon pair-production
threshold.  The following two first-order processes correspond to an
on-shell second-order process:  first the photon pair-produces, then
the initial electron annihilates with the newly produced positron to
produce the final photon [see Fig.\ \ref{nowidth}(a)].  If the
intermediate on-shell positron is in the Landau ground state, its decay
rate is zero, so that there is no line width supplied by the propagator
to control the resulting divergence.  This divergence represents
something of a calamity, since its effect is to make the total
cross-section for Compton scattering divergent everywhere above the
one photon pair-production threshold \cite{h79,dh86}.

A second example is provided by electron-electron scattering, in which
at least one of the initial electrons is in an excited Landau state
[Fig.\ \ref{nowidth}(b)].  Once again, there is a second-order on-shell
process that is analogous to a succession of two first order processes,
in which the excited electron emits a cyclotron photon which is in turn
absorbed by the other electron.  If the on-shell photon is below the
one photon pair-production threshold, its decay rate is zero, and there
is a resonant divergence --- see the expressions in Appendix C of
Ref.\ \cite{l81}.

It turns out that in every second-order process with a stable KAOSIS
there are non-zero decay rates associated with the initial
and final scattering states.  Thus, in the first of the two examples above,
the initial and final photons are capable of decay, while in the second
example, there are excited Landau levels in both the initial and final
states.  One might hope, then, that the decay rates of the scattering
states might be pressed into service to control the resonant divergences
when the decay rate of the intermediate state is zero.  As we now demonstrate,
this is not only possible, it is a necessary feature of the same program
of radiative corrections that brought the decay rates into the propagators.
That program demands that we should apply radiative corrections to the external
lines, in addition to the propagators.

It might be objected at this point that loop corrections to external
lines can have no bearing on the problem, since the arbitrary constants
which arise during renormalization are in part fixed by the requirement
that the external lines should suffer no corrections, so that the scattering
states should continue to be represented by solutions of the ``free''
wave equation with the physical mass \cite{bs59}.

The answer to this objection is that in the present case, the limited
number of arbitrary constants is not sufficient to satisfy the physical
requirement above for all scattering states.  For example, in the case
of the electron field, we may {\em only} demand that the Landau ground
state $A_G$ propagate according to Eq.\ (\ref{ldeqn}).  Once we have
used up the relevant renormalization constants to ensure this
condition, the remaining excited Landau states must propagate according
to Eq.\ (\ref{dldeqn}).  In other words, in Eq.\ (\ref{dldeqn}) we may
set $\Sigma(p^0=\eta E_G,A_G,\eta)=0$ (where $E_G$ is the energy of the
ground state), but then the remaining $A\neq A_G$ will yield non-zero
on-shell, diagonal matrix elements of $\Sigma$.  Similarly for the
Maxwell field, we may only demand that $\Pi(\omega_k,{\bf k},j)=0$ in
the limit ${\bf k}\rightarrow 0$, so that only infinitely soft photons
see no refringence in the magnetized vacuum.

The origin of this ``feature'' of magnetic QED is the fact that excited
scattering states are technically not scattering states at all, insofar
as they do not correspond to asymptotic one-particle states of the
quantum field which are stable.  In principle, we should only use
stable states as initial and final scattering states --- electrons and
positrons in the Landau ground state and photons below the one photon
pair-production threshold.  The consequence of this restriction on the
theory would be that multiple scattering events and scatterings
followed by multiple emissions could only be treated by computing
scattering amplitudes corresponding to very high order Feynman
diagrams, a notoriously burdensome task.  Thus, some of the most
interesting astrophysical applications of the theory would become
virtually inaccessible.

As an alternative, we may represent such high-order multiple events as
a succession of lower order transitions between states that are not
necessarily stable, and treating those states as if they were genuine
scattering states.  For example, a process in which an electron in the
ground state and a photon of energy above the third cyclotron
harmonic make a resonant transition to a state with an electron in the
ground state and four photons may be approximated
by a Compton scattering event in which the final electron is
left at the third harmonic, followed by three resonant decays.

This approximation of scattering states by excited states is a common
one in the literature \cite{dh86,bam86,hd91,km86}, but to date there
has been no investigation of its validity and limitations, or of what
modifications the usual Feynman perturbation theory must suffer in
order to accommodate them.  That investigation is the central concern
of this work.

We now discuss the specific modifications to the scattering states due
to radiative corrections.  Consider a bare external electron line in a
Feynman diagram which is represented in the scattering amplitude by the
spinor $\Psi^{(\eta)}_A(x)$, a solution of Eq.\ (\ref{ldeqn}).  After
subjecting the line to corrections associated with the self-energy
operator $\Sigma$, the result is a dressed line represented by the
spinor $\Theta^{(\eta)}_A(x)$, where
\begin{eqnarray}
\Theta^{(\eta)}_A&=&\Psi^{(\eta)}_A+G_B\cdot\Sigma\cdot\Psi^{(\eta)}_A
+G_B\cdot\Sigma\cdot
G_B\cdot\Sigma\cdot\Psi^{(\eta)}_A+\ldots\nonumber\\
&=&\Psi^{(\eta)}_A+G_B\cdot\Sigma\cdot\Theta^{(\eta)}_A.\label{dste1}
\end{eqnarray}
If we view the L-D operator [the wave operator in
Eq.\ (\ref{ldeqn})] as a ``free Hamiltonian'' and $\Sigma$ as a
perturbation, Eq.\ (\ref{dste1}) may be cast as a four-dimensional
Lippmann-Schwinger equation connecting an eigenstate $\Psi^{(\eta)}_A$
of the L-D operator with eigenvalue zero, to an eigenstate
$\Theta^{(\eta)}_A$ of the dressed L-D operator [the wave operator in
Eq.\ (\ref{dldeqn})], also with eigenvalue zero.  In other words,
$\Theta^{(\eta)}_A$ satisfies
\begin{equation}
[\gamma^\mu(i\partial_\mu-eA_\mu)-m-\Sigma]\cdot\Theta^{(\eta)}_A=0.
\label{dldeqn2}
\end{equation}
It is a simple matter to find solutions of this equation, since
we already know of states that simultaneously diagonalize the ``free''
operator and its perturbation.  Substituting
$\Theta^{(\eta)}_{A}(x)=e^{-ip^0x^0}\phi^{(\eta)}_A({\bf x})$ in
Eq.\ (\ref{dldeqn2}), we find
\begin{equation}
p^0=\eta(E_A-i\Gamma_A/2),\label{ecmpest}
\end{equation}
so that the dressed scattering state is
\begin{equation}
\Theta^{(\eta)}_{A}(x)=e^{-i\eta(E_A-i\Gamma_A/2)x^0}\phi^{(\eta)}_A({\bf x}).
\label{dest1}
\end{equation}

We may repeat the above argument for a bare external line which is represented
by the Dirac conjugate spinor $\overline{\Psi^{(\eta)}_A}(x)$.  The result
is that the dressed state is $\overline{\Lambda^{(\eta)}_A}(x)$, where
\begin{equation}
\overline{\Lambda^{(\eta)}_A}(x)=e^{+i\eta(E_A-i\Gamma_A/2)x^0}
\overline{\phi^{(\eta)}_A}({\bf x}).\label{dest2}
\end{equation}
Note that $\overline{\Lambda^{(\eta)}_A}(x)$ is {\em not} the Dirac
conjugate spinor of $\Theta^{(\eta)}_{A}(x)$, since the real part of
the exponent changes sign.

The procedure is analogous for the external photon lines.  The dressed
states $a(x)^\nu$ are solutions of
\begin{equation}
(g_{\mu\nu}\square-\Pi_{\mu\nu})a^\nu=0.\label{dmeqn}
\end{equation}
Using the polarization states of the previous section (for $j=\perp,\|$)
to write
\begin{equation}
{a_{{\bf k}j}}(x)^\nu=(2\omega_kL^3)^{-1/2}(\epsilon_j)^\nu
e^{-ik^0x^0+i{\bf k}\cdot{\bf x}},
\end{equation}
and substituting in Eq.\ (\ref{dmeqn}), we obtain
\begin{equation}
k^0=\pm[\omega_k-i\Gamma({\bf k},j)/2],\label{ecmppst}
\end{equation}
so that the dressed scattering state is
\begin{equation}
{a_{{\bf k}j}}(x)^\nu=(2\omega_kL^3)^{-1/2}(\epsilon_j)^\nu
e^{\pm i[\omega_k-i\Gamma({\bf k},j)/2]x^0-i{\bf k}\cdot{\bf x}}
.\label{dpst}
\end{equation}

{}From Eqs.\ (\ref{dest1}), (\ref{dest2}), and (\ref{dpst}) it is
apparent that the prescription of Eq. (\ref{ecmplx}) applies equally
well to scattering states as to propagators.  It is a general feature
of this prescription that the resulting positive and negative energy
states are not conjugate to each other, since the real part of the
exponent changes sign.  Consequently, positive energy solutions decay
as time increases, while negative energy solutions grow.  This is in
keeping with the interpretation of the negative energy solutions as
particles which move backwards in time.  In the application of these
formulae to the calculation of S-matrix elements, the
$e^{-i(E-i\Gamma/2)x^0}$ dependence is ascribed to initial states,
while the $e^{+i(E-i\Gamma/2)x^0}$ dependence is ascribed to final
states.

There is a second way of understanding the introduction of decay rates
in the time-development exponentials of scattering states.  We may take
the view, discussed above, that the metastable scattering states are
really approximations standing in for internal lines of much larger
Feynman diagrams.  That being the case, their functional form may be
read directly from the components of their respective propagators,
written in the forms of Eqs. (\ref{depropb}) and (\ref{dppropb}).

The appearance of the decay rates in the time-development exponentials
of the states will ultimately lead to their appearance in energy
denominators of scattering amplitudes, after integration over time
variables.  Here, however, there appears a major difference with the
usual practice of obtaining amplitudes.  The real parts of the
exponentials will lead in general to divergent expressions if the time
integration limits are allowed to go to $\pm\infty$ as usual.  It is
not difficult to see physically why we should expect trouble in this
limit.  Letting the upper time limit go to infinity in the S-matrix
element is tantamount to asking the question, ``what is the probability
of observing this decaying state in the infinitely distant future?''
Clearly, no calculation is required to see that the answer must be
``zero.''  Similarly, letting the lower time limit go to minus infinity
amounts to inquiring about an interaction at a finite time of a
decaying state which was prepared in the infinitely distant past --- a
process which also has vanishing probability of occurrence.

It is therefore necessary that the scattering theory be formulated
for states prepared at finite times $T_i$ and measured at finite
times $T_f$.  Only when the initial state is stable may the limit
$T_i\rightarrow -\infty$ be taken, and only when the final, measured
state is stable may we set $T_f\rightarrow\infty$.  These limits
are termed the ``absorption'' and ``emission'' limits, for reasons
that will shortly be made clear.

Note that due to the real parts in the time-development exponentials,
the scattering states are not generally normalized to unit probability.
In fact, they may only be so normalized at a given, fixed time.
Physically, it is necessary to ensure that the initial states be
normalized at the preparation time $T_i$, and that the final states
be normalized at the measurement time $T_f$.  This could be accomplished
by setting $x^0\rightarrow x^0-T_i$ or $x^0\rightarrow x^0-T_f$, as
appropriate, in the expressions of Eqs.\ (\ref{dest1}), (\ref{dest2}),
and (\ref{dpst}).  It is often simpler to calculate the amplitudes
with the states as written above and multiply the result by the factor
$\exp[\frac{1}{2}(\Gamma_iT_i-\Gamma_fT_f)]$, where $\Gamma_{i(f)}$
is the sum of the decay rates of the particles in the initial (final)
state.

There is an extremely important consequence of this finite-time
formulation of the theory: {\em strict energy conservation no longer
holds}.  The energy conserving $\delta$-functions which appeared in the
old amplitudes were a consequence of the integration of time
exponentials over all time.  By the time-energy uncertainty relation,
we may not determine the energy to infinite precision over a finite
time interval.  This is not an alarming consequence of the theory, but
rather a desirable one.  When we discuss the excitation or decay of
metastable states, we cannot expect to determine the energy of those
states to better than the natural line width, so energy-conserving
$\delta$-functions should actually violate our physical intuition for
these processes.  In fact, we will see that strict energy conservation
is recovered {\em only} when $\Gamma_i=\Gamma_f=0$.  Thus, for example,
in the expressions of Herold \cite{h79} for Compton scattering from
ground state to ground state, the energy-conserving $\delta$-functions
are appropriate.

We now discuss an example which provides a simple application of these
ideas: cyclotron decay.  Consider an electron which is prepared in an
excited state $A$ at a time $T_i=0$.  We calculate, to first order, the
probability that the system should make a transition to the ground
state $A_G$ with the emission of a photon in the state (${\bf k},j$),
which is below the one photon pair-production threshold.  Since
$\Gamma_f=0$, we may take the limit $T_f\rightarrow\infty$.  The
S-matrix element is given by
\begin{eqnarray}
S_{fi}&=&T_{A_GA}({\bf k})_\mu(\epsilon_j)^\mu
{\displaystyle\int_0^{\infty}}dx^0\,
e^{-i[(E_A-i\Gamma_A/2)-E_G-\omega_k]x^0}\nonumber\\
&=&\frac{-iT_{A_GA}({\bf k})_\mu(\epsilon_j)^\mu}
{E_A-E_G-\omega_k-i\Gamma_A/2},\label{em1}
\end{eqnarray}
where $T_{A_GA}({\bf k})_\mu$ is the interaction matrix element for the
transition:
\begin{equation}
T_{A_GA}({\bf k})^\mu=e^2L^{-3/2}(2\omega_k)^{-1/2}
\int d^3{\bf x}\,e^{-i{\bf k}\cdot{\bf x}}\,
\overline{\phi^{(+)}_{A_G}}({\bf x})
\gamma^\mu\phi^{(+)}_{A}({\bf x}).\label{emmel}
\end{equation}
Note that in Eq. (\ref{em1}), the energy conserving $\delta$-function
has been replaced by the Wigner-Weisskopf line shape function.  Thus,
the new formalism has reproduced the well-known result from
non-relativistic quantum mechanics, which the old formalism could not
(compare Refs.\ \cite{mz81,hrw82,k54,dh83,t52}).

We close this section with a discussion of the effect of using the
dressed electron scattering states of Eqs.\ (\ref{dest1}) and
(\ref{dest2}) on the range of the interaction in the presence of a
KAOSIS of the photon propagator below the one photon pair-production
threshold.  We may attempt to repeat the procedure that led to
Eqs.\ (\ref{effrange1}) and (\ref{effrange2}).  However, this time the
computation of the S-matrix element corresponding to
Fig.\ \ref{e+e-}(a) is no longer equivalent to taking the Fourier
transform of the photon propagator with respect to $x^0$, since the
dependence of the scattering states on $x^0$ is no longer purely
oscillatory.  Rather, the effective one-dimensional interaction
$g(x^1;k^2,k^3)$ that results is given by
\begin{equation}
g(x^1;k^2,k^3)=\int\frac{dk^0}{2\pi}\,f_B(x^1;k^0,k^2,k^3)
W(k^0-E),\label{effrange3}
\end{equation}
where $W(z)$ is a function which is only appreciable in a range
$\pm\Delta$ about $z=0$.  The restricted domain of $W$ is a reflection
of the restricted domain in time of the scattering states --- in fact,
we have either $\Delta\sim\Gamma$ or $\Delta\sim (T_f-T_i)^{-1}$,
whichever is largest.

By using stationary phase arguments, it is easy to see that $g(x^1;k^2,k^3)$
can only be appreciable for a limited range of $|x^1|$:
\begin{equation}
|x^1|\lesssim\frac{2\pi}{\left[(E+\Delta)^2-(k^2)^2-(k^3)^2\right]^{1/2}
-\left[E^2-(k^2)^2-(k^3)^2\right]^{1/2}}.\label{effrange4}
\end{equation}
Thus the range of the interaction is curtailed when the scattering states
may decay.  We see that the spurious infinite interaction range
that entered calculations which used bare excited scattering states
was a consequence of their assumed infinite duration.  Once
their limited duration is incorporated into the formalism, their interaction
becomes short-ranged.

Note that a KAOSIS of the photon propagator may exist only if either
some of the scattering states are excited or the KAOSIS itself is above
the one photon pair-production threshold.  Therefore, the scattering
cross-sections for processes with virtual photons {\em are now always
finite} if the dressed states and propagators are used to calculate
them.  We will show in the next section that resonant divergences are
now also under control in processes with virtual fermions.

\section{Scattering formulae}\label{sec_sf}
We now compute generic second-order formulae for two-particle to
two-particle scattering.  The Feynman diagram in Fig.\ \ref{genscat}
depicts such a process irrespective of whether the various lines
represent fermions or photons.  Our notation, which is illustrated in
Fig.\ \ref{genscat}, is as follows.  Let the energies of the external
lines be $E_\rho$, and let their decay rates be $\Gamma_\rho$
($\rho=a,b,c,d$).  Define $\mu_\rho=\pm 1$, with $\mu_\rho=+1$ if the
line represents an incoming state and $\mu_\rho=-1$ if it represents an
outgoing state.  Let the lines with $\rho=a,b$ join at the vertex with
coordinates $x$, and those with $\rho=c,d$ join at the vertex with
coordinates $y$ (see Fig.\ \ref{genscat}).  Define
$E_x\equiv\mu_aE_a+\mu_bE_b$,
$\Gamma_x\equiv\mu_a\Gamma_a+\mu_b\Gamma_b$, and similarly for $E_y$
and $\Gamma_y$.  Also define the complex energies ${\cal E}_x\equiv
E_x-\frac{i}{2}\Gamma_x$, ${\cal E}_y\equiv E_y-\frac{i}{2}\Gamma_y$.

Let the energies of the two initial particles be $e_{i1}$, $e_{i2}$,
and let their decay rates be $\gamma_{i1}$, $\gamma_{i2}$.  Also let
the energies of the two final particles be $e_{f1}$, $e_{f2}$,
and let their decay rates be $\gamma_{f1}$, $\gamma_{f2}$.  The $e$
and $\gamma$ are set equal to the $E$ and $\Gamma$ as appropriate to
the process under consideration, an identification illustrated by the passage
from Fig.\ \ref{genscat} to Fig.\ \ref{gendc}.

We denote the quantum state of the intermediate particle by $l$:  $l=A$
if the particle is a fermion, $l=({\bf k},j)$ if it is a photon.  The
energy and decay rate of the intermediate state are $E_l$ and
$\Gamma_l$, respectively, and we define ${\cal E}_l\equiv
E_l-\frac{i}{2}\Gamma_l$.  We also introduce the index $\eta=\pm 1$,
where $\eta=+1$ if the intermediate state has positive energy and
$\eta=-1$ otherwise.

The propagators are given by Eqs.\ (\ref{depropb}) and (\ref{dppropb}).
The S-matrix element for the process is obtained by sandwiching
the appropriate propagator between the scattering states in the usual
way \cite{bs59} and integrating over the space time coordinates $x$ and $y$.
The general result is
\begin{equation}
S_{fi}=i
{\displaystyle\sum_{l\eta}}c_\eta\, M_{fi}(l,\eta)\,
\xi^{(l\eta)}_{T_iT_f}({\cal E}_x,{\cal E}_y).\label{amp}
\end{equation}
Here, $M_{fi}(l,\eta)$ is the product of two interaction matrix
elements appropriate to the process, $c_\eta$ is 1 if the intermediate
state is a photon and $-\eta$ if it is a fermion, and
\begin{eqnarray}
\xi^{(l\eta)}_{T_iT_f}({\cal E}_x,{\cal E}_y)&\equiv&
e^{+(\gamma_{i1}+\gamma_{i2})T_i/2-(\gamma_{f1}+\gamma_{f2})T_f/2}
e^{+i(e_{i1}+e_{i2})T_i-i(e_{f1}+e_{f2})T_f}\nonumber\\
&&\times{\displaystyle\int_{T_i}^{T_f}dx^0\,\int_{T_i}^{T_f}dy^0}\,
\theta\left[\eta\,(x^0-y^0)\right]
\exp\left[-i\eta{\cal E}_l(x^0-y^0)-i{\cal E}_xx^0-i{\cal E}_yy^0\right].
\label{etfac1}
\end{eqnarray}
The factor
$e^{+(\gamma_{i1}+\gamma_{i2})T_i/2-(\gamma_{f1}+\gamma_{f2})T_f/2}$
in Eq.\ (\ref{etfac1}) adjusts the normalization of the initial and
final states to be 1 at $T_i$ and $T_f$, respectively, while the factor
$e^{+i(e_{i1}+e_{i2})T_i-i(e_{f1}+e_{f2})T_f}$ allows a convenient choice
of phase.  The quantity $\xi^{(l\eta)}_{T_iT_f}({\cal E}_x,{\cal E}_y)$
is the new object which incorporates the resonant energy denominators
and in general replaces the energy-conserving $\delta$-function.  It
may be calculated by the substitution $u=x^0-y^0$, $v=x^0+y^0$.  The
result is
\begin{eqnarray}
\xi^{(l\eta)}_{T_iT_f}({\cal E}_x,{\cal E}_y)&=&\quad
\frac{e^{+(\gamma_{i1}+\gamma_{i2})T_i/2-(\gamma_{f1}+\gamma_{f2})T_f/2}
e^{+i(e_{i1}+e_{i2})T_i-i(e_{f1}+e_{f2})T_f}}
{{\cal E}_x+{\cal E}_y}\nonumber\\
&&\times\left\{\quad
\frac{
e^{-i({\cal E}_x+{\cal E}_y)T_f}
\left[e^{(i/2)[(1-\eta){\cal E}_x+(1+\eta){\cal E}_y-2{\cal E}_l](T_f-T_i)}-1
\right]}
{(1/2)[(1-\eta){\cal E}_x+(1+\eta){\cal E}_y-2{\cal E}_l]}\right.\nonumber\\
&&\left.\quad-\quad\frac{
e^{-i({\cal E}_x+{\cal E}_y)T_i}
\left[e^{(i/2)[(-1-\eta){\cal E}_x+(-1+\eta){\cal E}_y-2{\cal E}_l](T_f-T_i)}-1
\right]}
{(1/2)[(-1-\eta){\cal E}_x+(-1+\eta){\cal E}_y-2{\cal E}_l]}\right\}.
\label{etfacgen}
\end{eqnarray}

Eq.\ (\ref{etfacgen}) provides an expression which may be adapted to
any second-order process of interest by adjusting the $\mu$ and the
assignments of the $e$ and $\gamma$ to the $E$ and $\Gamma$.  In the
particular case of two-particle to two-particle scattering we need
consider four cases: the ``direct'' and ``cross'' diagrams
(Fig.\ \ref{gendc}), each for $\eta=\pm1$.  The expressions are
simplified by the notation $\Delta e\equiv
e_{i1}+e_{i2}-e_{f1}-e_{f2}$,
$\Delta\gamma\equiv\gamma_{i1}+\gamma_{i2}-\gamma_{f1}-\gamma_{f2}$,
and $\tau\equiv T_f-T_i$.

\paragraph{Direct diagram, $\eta=+1$}\label{dplhd}
We obtain this case from Eq.\ (\ref{etfacgen}) by setting ${\cal
E}_x=-(e_{f1}+e_{f2})+i(\gamma_{f1}+\gamma_{f2})/2$ and ${\cal
E}_y=(e_{i1}+e_{i2})-i(\gamma_{i1}+\gamma_{i2})/2$.  We obtain
\begin{eqnarray}
\xi^{(l,+1)}_{T_iT_f}({\cal E}_x,{\cal E}_y)&=&
\frac{1}{\Delta e-i\Delta\gamma/2}
\times\left\{
\frac{e^{-i(E_l-i\Gamma_l/2)\tau}-
e^{-i[(e_{i1}+e_{i2})-i(\gamma_{i1}+\gamma_{i2})/2]\tau}}
{(e_{i1}+e_{i2}-E_l)-i(\gamma_{i1}+\gamma_{i2}-\Gamma_l)/2}\right.
\nonumber\\
&&\left.-\frac{e^{-i(E_l-i\Gamma_l/2)\tau}-
e^{-i[(e_{f1}+e_{f2})-i(\gamma_{f1}+\gamma_{f2})/2]\tau}}
{(e_{f1}+e_{f2}-E_l)-i(\gamma_{f1}+\gamma_{f2}-\Gamma_l)/2}\right\}.
\label{dpl}
\end{eqnarray}

\paragraph{Direct diagram, $\eta=-1$}\label{dmnhd}
For the same assignments as case (\ref{dplhd}), but with $\eta=-1$, we
find
\begin{eqnarray}
\xi^{(l,-1)}_{T_iT_f}&&({\cal E}_x,{\cal E}_y)=
\frac{1}{\Delta e-i\Delta\gamma/2}\nonumber\\
&&\times\left\{
\frac{e^{-i[(e_{i1}+e_{i2}+e_{f1}+e_{f2}+E_l)
-i(\gamma_{i1}+\gamma_{i2}+\gamma_{f1}+\gamma_{f2}+\Gamma_l)/2]\tau}
-e^{-i[(e_{i1}+e_{i2})-i(\gamma_{i1}+\gamma_{i2})/2]\tau}}
{(-e_{f1}-e_{f2}-E_l)-i(-\gamma_{f1}-\gamma_{f2}-\Gamma_l)/2}\right.
\nonumber\\
&&\left.-
\frac{e^{-i[(e_{i1}+e_{i2}+e_{f1}+e_{f2}+E_l)
-i(\gamma_{i1}+\gamma_{i2}+\gamma_{f1}+\gamma_{f2}+\Gamma_l)/2]}
-e^{-i[(e_{f1}+e_{f2})-i(\gamma_{f1}+\gamma_{f2})/2]\tau}}
{(-e_{i1}-e_{i2}-E_l)-i(-\gamma_{i1}-\gamma_{i2}-\Gamma_l)/2}\right\}.
\label{dmn}
\end{eqnarray}

\paragraph{Cross diagram, $\eta=+1$}\label{cplhd}
Here we set ${\cal
E}_x=(e_{i2}-e_{f2})-i(\gamma_{i2}-\gamma_{f2})/2$ and
${\cal E}_y=(e_{i1}-e_{f1})-i(\gamma_{i1}-\gamma_{f1})/2$:
\begin{eqnarray}
\xi^{(l,+1)}_{T_iT_f}({\cal E}_x,{\cal E}_y)&=&
\frac{1}{\Delta e-i\Delta\gamma/2}\times\left\{
\frac{e^{-i[(e_{f1}+e_{i2}+E_l)-i(\gamma_{f1}+\gamma_{i2}+\Gamma_l)/2]\tau}-
e^{-i[(e_{i1}+e_{i2})-i(\gamma_{i1}+\gamma_{i2})/2]\tau}}
{(e_{i1}-e_{f1}-E_l)-i(\gamma_{i1}-\gamma_{f1}-\Gamma_l)/2}\right.\nonumber\\
&&\left.-\frac{
e^{-i[(e_{f1}+e_{i2}+E_l)-i(\gamma_{f1}+\gamma_{i2}+\Gamma_l)/2]\tau}-
e^{-i[(e_{f1}+e_{f2})-i(\gamma_{f1}+\gamma_{f2})]\tau}}
{(e_{f2}-e_{i2}-E_l)-i(\gamma_{f2}-\gamma_{i2}-\Gamma_l)/2}
\right\}.
\label{cpl}
\end{eqnarray}

\paragraph{Cross diagram, $\eta=-1$}\label{cmnhd}
The assignments here are as in case (\ref{cplhd}), and $\eta=-1$:
\begin{eqnarray}
\xi^{(l,-1)}_{T_iT_f}({\cal E}_x,{\cal E}_y)&=&
\frac{1}{\Delta e-i\Delta\gamma/2}\times\left\{
\frac{e^{-i[(e_{f2}+e_{i1}+E_l)-i(\gamma_{f2}+\gamma_{i1}+\Gamma_l)/2]\tau}-
e^{-i[(e_{i1}+e_{i2})-i(\gamma_{i1}+\gamma_{i2})/2]\tau}}
{(e_{i2}-e_{f2}-E_l)-i(\gamma_{i2}-\gamma_{f2}-\Gamma_l)/2}
\right.\nonumber\\
&&\left.-\frac{
e^{-i[(e_{f2}+e_{i1}+E_l)-i(\gamma_{f2}+\gamma_{i1}+\Gamma_l)/2]\tau}-
e^{-i[(e_{f1}+e_{f2}+E_l)-i(\gamma_{f1}+\gamma_{f2}+\Gamma_l)/2]\tau}}
{(e_{f1}-e_{i1}-E_l)-i(\gamma_{f1}-\gamma_{i1}-\Gamma_l)/2}\right\}.
\label{cmn}
\end{eqnarray}

The expressions in Eqs.\ (\ref{dpl}), (\ref{dmn}),
(\ref{cpl}), and (\ref{cmn}) all contain products of two complex energy
denominators, one ``energy-conserving'' and the other resonant.  All
these energy denominators may in principle go to zero.  When this happens,
however, there is always a cancellation of the exponentials in the numerators,
so that the result is always finite.  This is a consequence of the fact
that these expressions were derived starting from Eq.\ (\ref{etfac1}), an
integral over a finite range of a finite integrand, which consequently may
never diverge.  Therefore, scattering processes containing virtual
fermions now have finite S-matrix elements.  These were the last
remaining resonant divergences in the theory, {\em which is now entirely
finite}.

We have thus eliminated all resonant divergences from our scattering
cross-sections.  The price we have paid is the
dependence of the S-matrix elements on the time lapse $\tau$ between the
preparation of the initial state and the measurement of the final
state, and the attendant loss of strict energy conservation.  Note that
in general, the nature of the dependence on $\tau$ is for the
S-matrix elements to decay away as $\tau\rightarrow\infty$.  As discussed in
the previous section, this is the behavior expected for scattering from
excited states to excited states.

Note also that these expressions lack crossing symmetry.  The reason is the
introduction of the exponential factor outside the integral in
Eq.\ (\ref{etfacgen}), which is not symmetric under the replacement
$e_{i1}\leftrightarrow -e_{f2}$, $\gamma_{i1}\leftrightarrow
-\gamma_{f2}$.  If we divide the expressions of
Eqs.\ (\ref{dpl}), (\ref{dmn}), (\ref{cpl}), and (\ref{cmn}) by the
exponential factor we find that the resulting expressions are, in
fact, crossing symmetric.

In order to parlay the above expressions into cross-sections, we must
substitute them into Eq.\ (\ref{amp}) to obtain a $\tau$-dependent
expression for the S-matrix element.  The reaction rate $R$ is then
given by $R=d|S_{fi}|^2/d\tau=2\text{Re}\{{S_{fi}}^*dS_{fi}/d\tau\}$.
The cross-section may be obtained from $R$ by the usual kinematic
manipulation: $d\sigma/d\Omega_f=L^3|{\bf v}|^{-1}R$, where $d\Omega_f$
is a volume element in the space of final states and ${\bf v}$ is the
relative velocity of the initial particles.

There are two special cases where it is possible to eliminate the dependence
on $\tau$ from these expressions: when the initial state is stable
($\gamma_{i1}=\gamma_{i2}=0$), and when the final state is stable
($\gamma_{f1}=\gamma_{f2}=0$).  These cases are called ``absorption''
and ``emission'' scattering, respectively.  In the absorption scattering
case, we may set $T_i\rightarrow-\infty$, while in the emission
scattering case we may set $T_f\rightarrow\infty$.  In either case,
$\tau\rightarrow\infty$, and the above expressions for the $\xi$
are greatly simplified.

\subsection{Absorption scattering}

The special case of a stable initial state corresponds to a situation
analogous to absorption, in which we prepare beams of stable particles
and observe the excited products before they have the opportunity to
decay.  If we set $\gamma_{i1}=\gamma_{i2}=0$ in Eqs.\ (\ref{dpl}),
(\ref{dmn}), (\ref{cpl}), and (\ref{cmn}), and take the limit
$\tau\rightarrow\infty$, we find the following results:

For the direct diagram,
\begin{mathletters}
\label{absetfd}
\begin{equation}
\xi^{(l,+1)}_{-\infty,T_f}({\cal E}_x,{\cal E}_y)=
\frac{1}{(\Delta e-i\Delta\gamma/2)
[(-e_{i1}-e_{i2}+E_l)-i\Gamma_l/2]},\label{absetfdpl}
\end{equation}
\begin{equation}
\xi^{(l,-1)}_{-\infty,T_f}({\cal E}_x,{\cal E}_y)=
\frac{1}{(\Delta e-i\Delta\gamma/2)
[(e_{f1}+e_{f2}+E_l)-i(\gamma_{f1}+\gamma_{f2}+\Gamma_l)/2]},\label{absetfdmn}
\end{equation}
\end{mathletters}
and for the cross diagram,
\begin{mathletters}
\label{absetfc}
\begin{equation}
\xi^{(l,+1)}_{-\infty,T_f}({\cal E}_x,{\cal E}_y)=
\frac{1}{(\Delta e-i\Delta\gamma/2)
[(e_{f1}-e_{i1}+E_l)-i(\gamma_{f1}+\Gamma_l)/2]},\label{absetfcpl}
\end{equation}
\begin{equation}
\xi^{(l,-1)}_{-\infty,T_f}({\cal E}_x,{\cal E}_y)=
\frac{1}{(\Delta e-i\Delta\gamma/2)
[(e_{f2}-e_{i2}+E_l)-i(\gamma_{f2}+\Gamma_l)/2]}.\label{absetfcmn}
\end{equation}
\end{mathletters}
We have eliminated an inessential phase factor.  Note that
$\Delta\gamma=-(\gamma_{f1}+\gamma_{f2})$.  Substituting these
expressions in Eq.\ (\ref{amp}), we obtain a time-independent S-matrix
element.  In order to get the result into the form of a cross-section,
note that the final state is decaying at a rate
$\gamma_{f1}+\gamma_{f2}$.  In order that the probability of that
final state be time-independent and equal to $|S_{fi}|^2$, the reaction
rate must exactly balance the decay rate, so that we must have
$R=(\gamma_{f1}+\gamma_{f2})|S_{fi}|^2=|\Delta\gamma||S_{fi}|^2$.
Again, cross-section $d\sigma/d\Omega_f$ may be trivially obtained from
$R$.  Note that $d\sigma/d\Omega_f$ contains the following functional
dependence on $\Delta e$:
\begin{equation}
\frac{d\sigma}{d\Omega_f}\propto\frac{\left|\Delta\gamma\right|}
{\Delta e^2+(\Delta\gamma/2)^2}
\stackrel{\Delta\gamma\rightarrow 0}{\longrightarrow}2\pi\,
\delta(e_{i1}+e_{i2}-e_{f1}-e_{f2}).\label{absecon}
\end{equation}
In other words, the energy conservation in the
cross-section is Lorentzian, and in the limit of stable final states
we recover the energy-conserving $\delta$-function with the correct
coefficient of $2\pi$.  As expected, the strict energy conservation in
the old expressions for the S-matrix elements is correct for scattering
from stable states to stable states.

\subsection{Emission scattering}
The special case of a stable final state corresponds to a situation
analogous to emission, in which we prepare a beam of excited
particles and observe them after they have scattered into the stable
final state.  We set $\gamma_{f1}=\gamma_{f2}=0$ in Eqs.\ (\ref{dpl}),
(\ref{dmn}), (\ref{cpl}), and (\ref{cmn}), and take the limit
$\tau\rightarrow\infty$, to obtain:

For the direct diagram,
\begin{mathletters}
\label{emetfd}
\begin{equation}
\xi^{(l,+1)}_{T_i,\infty}({\cal E}_x,{\cal E}_y)=
\frac{1}{(\Delta e-i\Delta\gamma/2)[(e_{f1}+e_{f2}+E_l)-i\Gamma_l/2]},
\label{emetfdpl}
\end{equation}
\begin{equation}
\xi^{(l,-1)}_{T_i,\infty}({\cal E}_x,{\cal E}_y)=
\frac{1}{(\Delta e-i\Delta\gamma/2)
[(e_{i1}+e_{i2}+E_l)-i(\gamma_{i1}+\gamma_{i2}+\Gamma_l)/2]},
\label{emetfdmn}
\end{equation}
\end{mathletters}
and for the cross diagram,
\begin{mathletters}
\label{emetfc}
\begin{equation}
\xi^{(l,+1)}_{T_i,\infty}({\cal E}_x,{\cal E}_y)=
\frac{1}{(\Delta e-i\Delta\gamma/2)
[(e_{i2}-e_{f2}+E_l)-i(\gamma_{i2}+\Gamma_l)/2]},
\label{emetfcpl}
\end{equation}
\begin{equation}
\xi^{(l,-1)}_{T_i,\infty}({\cal E}_x,{\cal E}_y)=
\frac{1}{(\Delta e-i\Delta\gamma/2)
[(e_{i1}-e_{f1}+E_l)-i(\gamma_{i1}+\Gamma_l)/2]}.
\label{emetfcmn}
\end{equation}
\end{mathletters}
Once again, we have eliminated an inessential phase from the
amplitudes.  Now we have $\Delta\gamma=\gamma_{i1}+\gamma_{i2}$.
Substituting these expressions in Eq.\ (\ref{amp}), we again obtain a
time-independent S-matrix element.  We obtain a cross-section by the
argument illustrated in Fig.\ \ref{tube}, which depicts a semi-infinite tube of
cross-sectional area $d\sigma/d\Omega_f$, terminating at the position
of the target particle and extending in the direction of
$-{\bf v}$.  The total probability (per unit final phase-space volume)
of an interaction leading to a final state in $d\Omega_f$ is equal
to the integral over the interior of the tube of the probability
that each infinitesimal slice should contain the projectile particle
{\em and} that it should actually reach the target particle
in spite of the fact that the two-particle state is decaying away:
\begin{equation}
|S_{fi}|^2=L^{-3}{\displaystyle\int_0^\infty}|{\bf v}|dt\,
\frac{d\sigma}{d\Omega_f}e^{-(\gamma_{i1}+\gamma_{i2})t}
=\frac{|{\bf v}|}{\Delta\gamma\,L^3}\frac{d\sigma}{d\Omega_f},
\end{equation}
so that
\begin{equation}
\frac{d\sigma}{d\Omega_f}=L^3|{\bf v}|^{-1}\Delta\gamma\,|S_{fi}|^2.
\end{equation}

We again see the Lorentzian energy conservation of Eq.\ (\ref{absecon}),
so that in this case we also recover strict energy conservation for stable
state to stable state scattering.

\section{Discussion}\label{sec_disc}
The regime of validity of the ``emission'' and ``absorption''
scattering limits is obvious from their context.  On the other hand,
the general formula in Eq.\ (\ref{etfacgen}) requires some discussion.
As discussed previously, the formula never ``misbehaves'', in the sense
that it never yields a divergent result.  In fact, in general that
result tends to zero as $\tau\rightarrow\infty$.  While this is
physically sensible, it obviously makes the large time-lapse limit a
less than useful one.  It is clear that what fails in this limit is the
validity of the perturbation-theoretic order of the calculation.  Since
the scattering states are themselves decaying to other states, those
other states should be included in the calculation, leading to
higher-order processes.  The second-order calculations outlined in the
previous section are only useful for values of $\tau$ such that the
scattering states have little chance to decay, that is for
$\Gamma\tau\ll 1$, where $\Gamma$ is the largest of the decay rates in
the process.  For example, we might choose $\tau$ to be on the order of
a collision time, if we are studying a gas with density and temperature
such that the collision rates far exceed the decay rates.

The general case above obviously represents a fairly radical departure
from the usual scattering formulae.  The ``emission'' and
``absorption'' scattering limits amount to somewhat less radical
modifications.  One obvious such modification is the replacement of
strict energy conservation with ``Lorentzian'' energy conservation.
Another is that the ``non-resonant'' energy denominators
[Eqs.\ (\ref{absetfdmn}), (\ref{absetfcpl}), (\ref{absetfcmn}),
(\ref{emetfdmn}), (\ref{emetfcpl}), (\ref{emetfcmn})] now contain the
decay rates of the scattering states as well as those of the
intermediate states.  The importance of these changes depends upon
whether the decay widths entering the energy denominators are electron
decay widths or photon decay widths.

If the decay widths are purely fermionic, they are gently varying
functions of energy, and their magnitudes are smaller by $e^2$ than
their own characteristic scale of variation, the scale of variation of
the interaction matrix elements, and the characteristic separation of
the resonances.  Consequently, in this case the relative change that
results from introducing Lorentzian, rather than exact, energy
conservation, and from introducing the decay widths of the scattering
states into the energy denominators, is of order $e^2$.  On the other
hand, if some of the decay widths correspond to photon lines, they can
vary rather rapidly as a function of energy \cite{k54,dh83}.  Thus, the
behavior of the energy denominators is not really ``Lorentzian'',
despite notational appearances to the contrary.  The departure from the
cross-sections computed assuming strict energy conservation and not
including external line decay widths might turn out to be appreciable
in this case, although its precise magnitude remains to be assessed.

In the limit of stable scattering states the usual results are
completely reproduced, since we recover strict energy conservation
and there are no scattering state decay widths to include in resonant
energy denominators.

In connection with the dressed photon propagator of
Eq.\ (\ref{dppropa}), we wish to comment on a point which is a
potential source of confusion.  The decay width $\Gamma({\bf k},j)$ is
to be evaluated on the light cone, as implied by the first line of
Eq.\ (\ref{gammaphot}).  Now, when the photon propagator is used in an
S-matrix element, the values of $k^0$, $k^2$, and $k^3$ are fixed by
the $x^0$, $x^2$, and $x^3$ momenta of the scattering states.  Thus,
the sum over intermediate photon wave states involves an integral over
the component $k^1$ of the wave vector ${\bf k}$.  The value of
$\Gamma({\bf k},j)$ must be evaluated for {\em each} value of $k^1$ in the
integral.  This is analogous to the case of the electron propagator,
Eq.\ (\ref{depropa}), in which the intermediate electron decay width is
evaluated, on the energy shell, for each Landau level in the sum over
intermediate states.  The only difference between the two cases is that
the relevant degrees of freedom are discrete for the electron
propagator, while they are continuous for the photon propagator.

The radiatively corrected photon propagator permits for the first time
the evaluation of processes such as
$\text{e}^+\text{e}^-\rightarrow\text{e}^+\text{e}^-$,
$\text{e}^-\rightarrow\text{e}^-\text{e}^+\text{e}^-$, and
$\gamma\text{e}^-\rightarrow\text{e}^-\text{e}^+\text{e}^-$, all of
which are important for neutron star emission.  Of equal astrophysical
importance is the evaluation above the one photon pair-production
threshold of Compton scattering, two photon pair annihilation, and
two photon pair production, which is now possible by virtue of the
radiatively corrected scattering states.  Finally, we now have access
to the processes $\text{e}^-\text{e}^-\rightarrow\text{e}^-\text{e}^-$ and
$\text{e}^+\text{e}^-\rightarrow\text{e}^+\text{e}^-$ even when
the initial and final states are excited.

\section{Acknowledgments}
While this work was performed, Carlo Graziani held a National Research
Council-NASA Goddard Space Flight Center Research Associateship.

\begin{figure}
\caption{Two diagrams for e$^+$-e$^-$ scattering.  The labels $a_i^+$,
$a_i^-$ denote the $x^1$-coordinate of the orbit center of the initial
positron and electron, respectively, while the labels $a_f^+$,
$a_f^-$ denote the orbit center of the final positron and
electron, respectively.  The spatial separation between the two fermion
lines is given by the parameter $s=a_f^+-a_f^-$ for (a), and
by $t=a_f^--a_i^-$ for (b).  In either case the interaction does not
fall off with increasing separation, so that the result is a divergence
in the total cross-section.}
\label{e+e-}
\end{figure}
\begin{figure}
\caption{Processes allowing non-decaying, on-shell virtual states; (a)
$\text{e}^-\gamma$ scattering; (b) $\text{e}^-\text{e}^-$ scattering.
The process in (a) may be viewed as a pair production followed by a
pair annihilation.  When the intermediate positron is in the Landau
ground state it has zero decay width.  Similarly, the process in (b)
may be viewed as cyclotron emission followed by cyclotron absorption.
The decay width of the intermediate state vanishes if it is below the
one photon pair-production threshold.}
\label{nowidth}
\end{figure}
\begin{figure}
\caption{Generic scattering diagram.  The lines may be either fermion or
photon lines, and the external lines may be either incoming ($\mu=+1$)
or outgoing ($\mu=-1$).
}
\label{genscat}
\end{figure}
\begin{figure}
\caption{Generic ``direct'' and ``cross'' diagrams.  The lines may be
either fermion or photon lines.}
\label{gendc}
\end{figure}
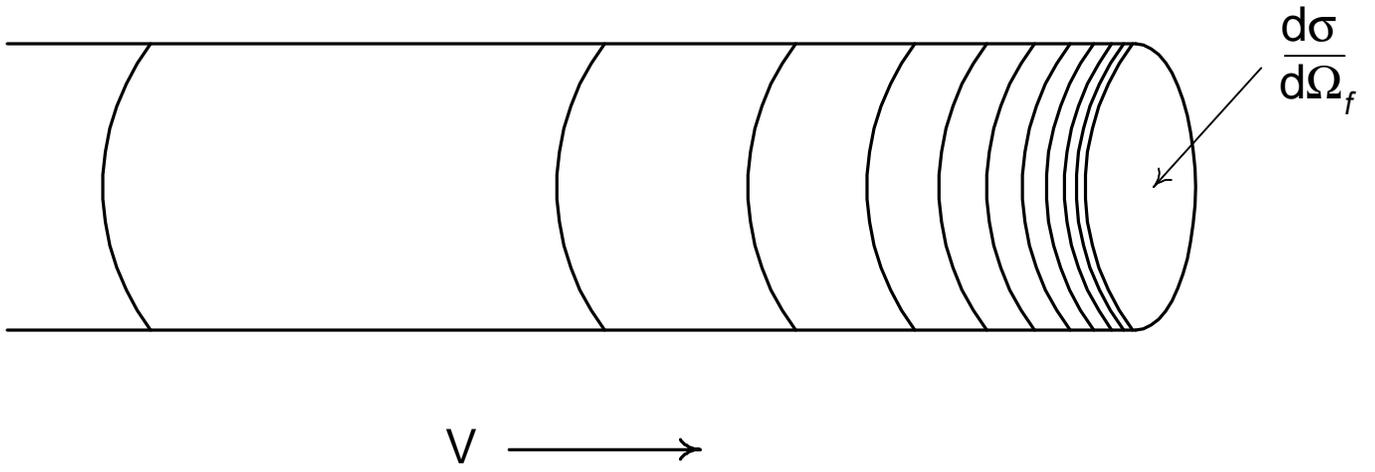
\begin{figure}
\caption{Kinematics of emission scattering.  The target particle is at
the end face of the semi-infinite tube of cross-sectional area
$d\sigma/d\Omega_f$.  The projectile particle must be in the tube in
order for an interaction to occur.  The decreasing density of circular
sections in the figure is meant to represent the decreasing probability
of an interaction due to the exponential decay of the initial state.}
\label{tube}
\end{figure}

\end{document}